# Enabling Skip Graphs to Process *K*-Dimensional Range Queries in a Mobile Sensor Network

Gregory J. Brault[1] Christopher J. Augeri Barry E. Mullins
Rusty O. Baldwin Christopher B. Mayer

*Department of Electrical and Computer Engineering*
*Air Force Institute of Technology, Wright-Patterson Air Force Base, Dayton, Ohio*
*Gregory.Brault@lackland.af.mil*
{*Chris.Augeri, Barry.Mullins, Rusty.Baldwin, Chris.Mayer*}*@afit.edu*

## Abstract

*A skip graph is a resilient application-layer routing structure that supports range queries of distributed k-dimensional data. By sorting deterministic keys into groups based on locally computed random membership vectors, nodes in a standard skip graph can optimize range query performance in mobile networks such as unmanned aerial vehicle swarms.*

*We propose a skip graph extension that inverts the key and membership vector roles and bases group membership on deterministic vectors derived from the z-ordering of k-dimensional data and sorting within groups is based on locally computed random keys.*

## 1. Introduction

Considerable research efforts are devoted to distributed sensor network (DSN) technologies which consist of devices with limited power and processing resources, but that are able to communicate with one another to execute user requests. In many instances, it is not feasible to conduct research on an actual implementation of a large-scale sensor network, e.g., an unmanned aerial vehicle (UAV) swarm containing 10,000 nodes. Thus, UAV swarm research necessarily uses theoretical analysis and simulation.

[1] Work performed while at the Air Force Institute of Technology. This work is supported in part by the Air Force Communications Agency. The views expressed in this paper are those of the authors and do not reflect the official policy or position of the United States Air Force, Department of Defense, or the U.S. Government.

Our research focuses on an application-layer routing structure called a skip graph which is used to execute queries of *k*-dimensional data that may be distributed across thousands of nodes in the network. The skip graph, herein referred to as a *standard* skip graph, was independently suggested in [1] and [6] as an extension to the skip list [9] for storing an arbitrary set of 1-dimensional (1-D) keys.

A skip graph member is associated with a *key* that users may wish to query, such as temperature or pressure sensor readings. Some *multi-dimensional* skip graph variants [3][5] have also been proposed to query *k*-dimensional (*k*-D) data, such as a sensor node's geographic coordinates. These extensions manipulate sorted keys in the skip graph versus the membership vectors that group the keys.

We propose an extension to the skip graph which increases query performance on *k*-D data distributed throughout the network in some querying conditions. Our goal is to minimize query response time in the application layer, not necessarily energy expended at each node as the response is to be time-critical. We also discuss the standard skip graph components that are modified in order to create our proposed multi-dimensional skip graph.

## 2. Linearizing *k*-dimensional data

Regardless of whether the data is 1-D or *k*-D, both the standard and multi-dimensional skip graphs store the data as 1-D keys. Therefore, *k*-D data must first be *linearized* to a single dimension. Figure 1(a) illustrates a common technique used for reducing dimensionality, the *z*-order space-filling curve [3], whose name is derived from the "*Z*"-like pattern formed as it snakes through the *k*-D area of interest. The *z*-curve has been independently discovered in many contexts [10] and is also known as Morton Ordering [7]. The distributed and dynamic nature of our application precludes the use of more optimal linearization methods, e.g., multi-dimensional scaling [2] or PageRank [8].

Figure 1(b) illustrates a grid populated with eight nodes represented as circles and labeled with a node identification number. The number inside each node is a sensor reading that will be used in following query discussions. The numbers outside the circles represent the node identification. The *x* and *y* axes in Figure 1(b) are represented in binary to better illustrate the *z*-order bit interleaving process.

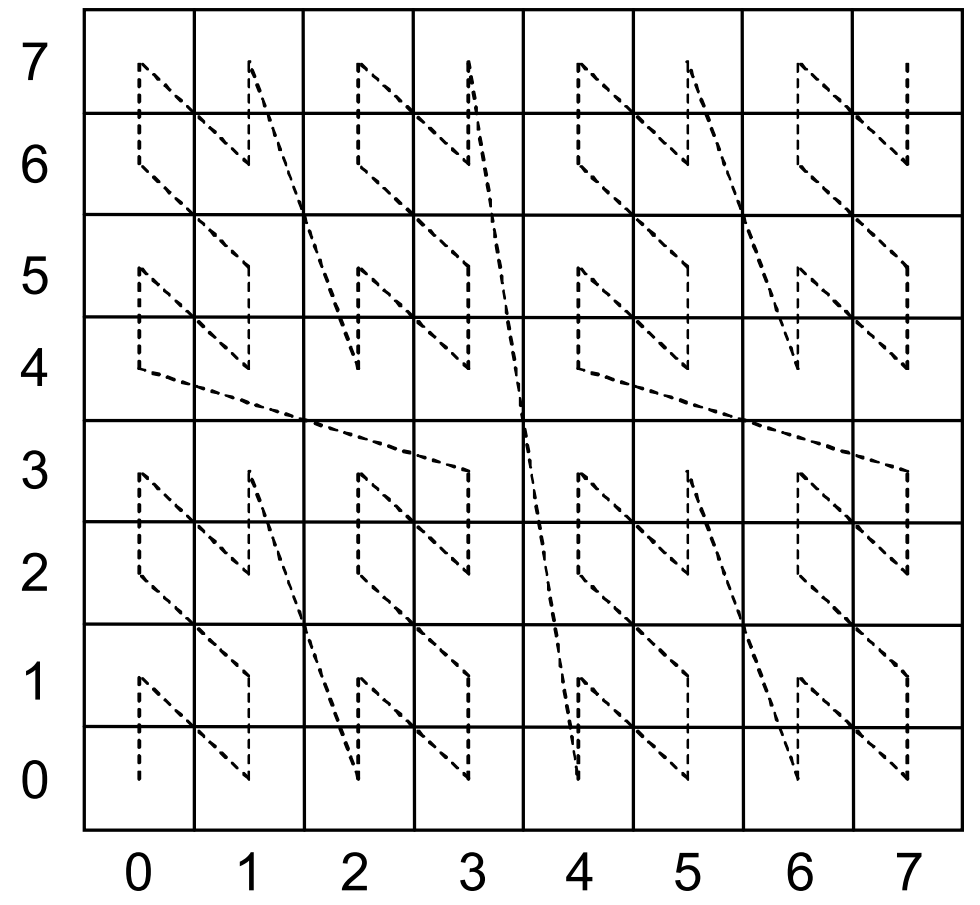

(a) 2-D z-order curve

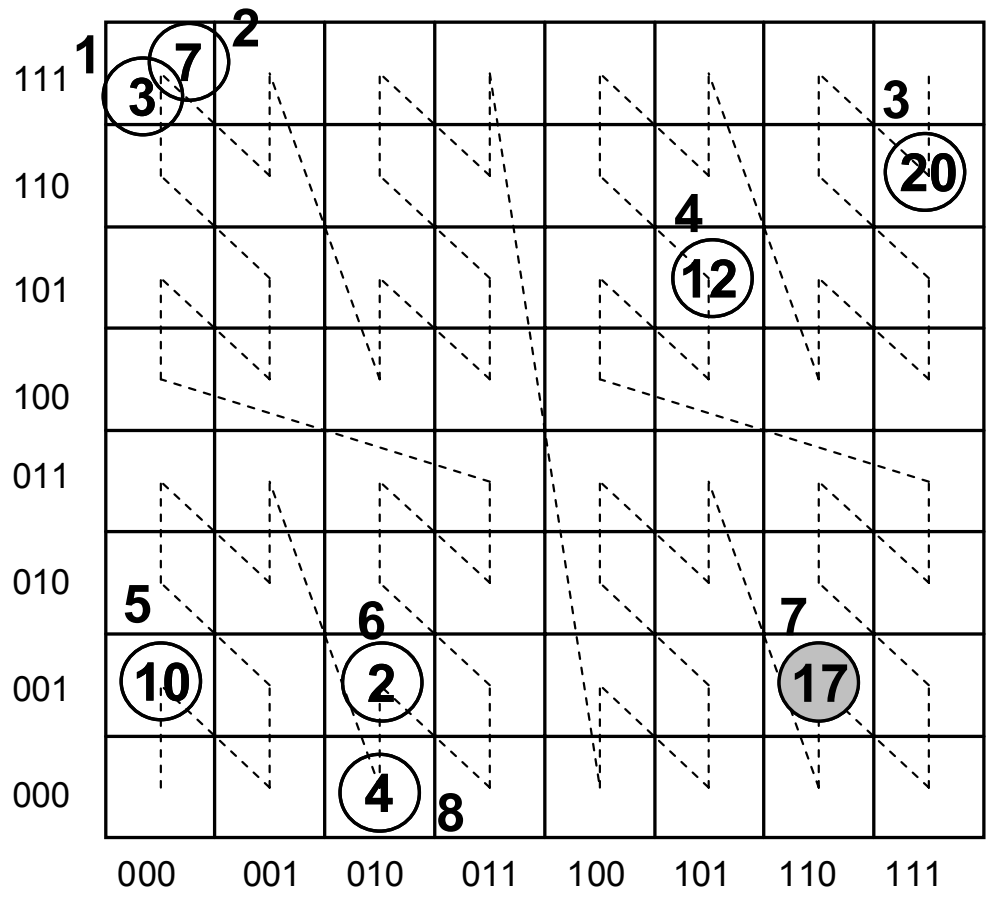

(b) Geographic area populated with nodes

**Figure 1. 2-D *Z*-Order Curve Examples**

Several properties of a *z*-order curve make it useful for querying geographical data. First, *z*-ordering is logically equivalent in two dimensions to a quad-tree [4][11][12][7], where a curve divides 2-D space into four equal-sized quadrants that are recursively defined using *z*-ordering. Additionally, *z*-ordering is extremely useful in sensor networks, since a node can compute it efficiently and locally, i.e., the algorithm executes in a decentralized manner, a critical property in order to maintain scaling in a distributed sensor network.

The interleaving process is shown in Figure 2(a), where we construct a *z*-order key based on the coordinates of node ‘7’ from Figure 1. The bits of the *k* coordinates are simply “zipped” together and the resulting value serves as the 1-D key of node ‘7’. This process easily extends to an arbitrary number of *k* dimensions, as Figure 2(b) shows (the value of ‘5’ for the third dimension was chosen arbitrarily).

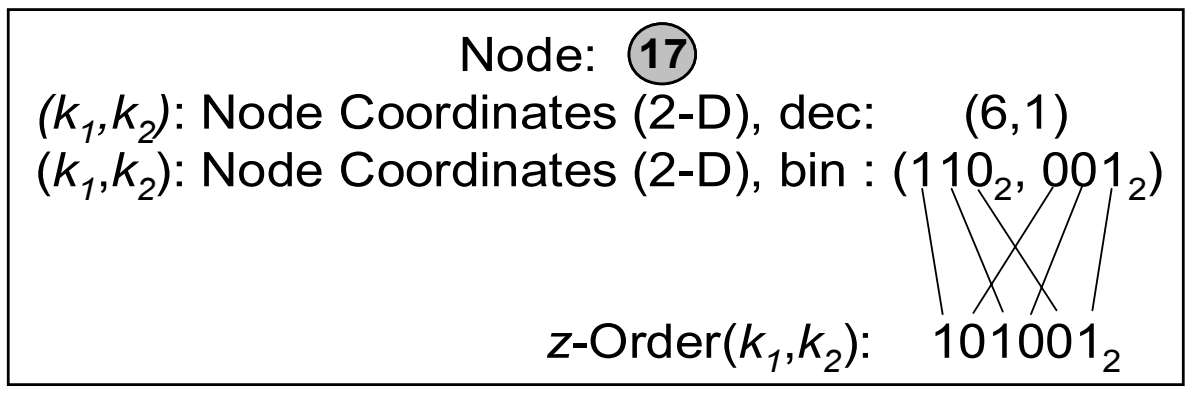

(a) Mapping 2-D coordinates

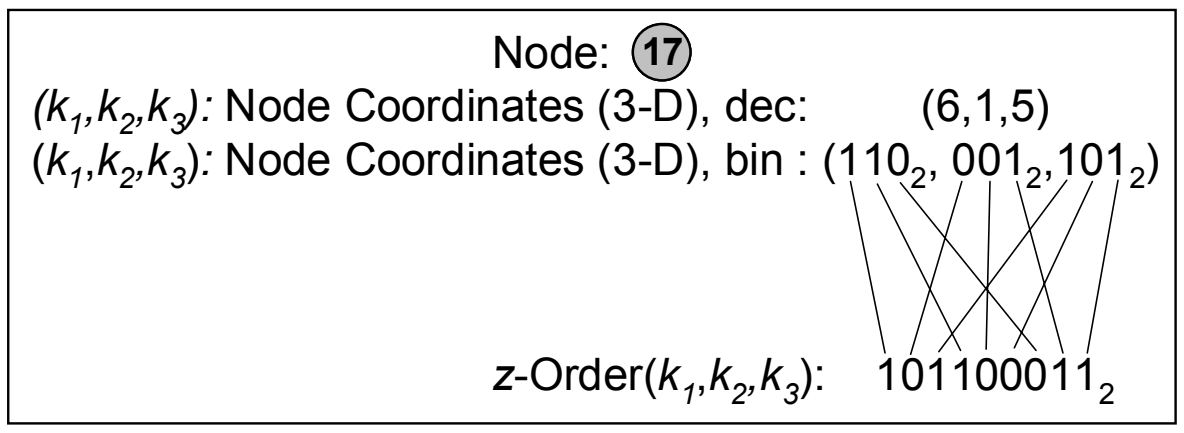

(b) Mapping 3-D coordinates

**Figure 2. Computing keys by *z*-ordering**

# 3. Standard uni-dimensional skip graph

## 3.1. Construction

A skip graph is a distributed data structure existing within a sensor network. There are multiple levels in a skip graph, and nodes are grouped into increasingly smaller lists within each successively higher level. Group membership is based on a membership vector, computed locally at each node. As a node is added to a higher level, it links itself into a list based on an increasingly longer prefix match of the nodes' membership vectors. Nodes are sorted within lists based on node key values, which are also computed locally within each node.

In a *uni-dimensional* standard skip graph, the node keys are determined by scalar data, e.g., temperature or pressure sensor readings, and the membership vectors are computed randomly. Figure 3 is an example of a uni-dimensional standard skip graph, where the circles represent nodes, and the numbers inside the circles indicate the node's key value—the single-dimensional data associated with the node. The binary sequence below each node indicates the node's membership vector, and the prefix of the membership vector being used to group the nodes into lists on a given level is underlined. The links between nodes represent logical communication links in the sensor network. The construction and query examples that follow are based on the sensor network illustrated in Figure 1. Nodes are referenced by key value, e.g., the node in Figure 1 with a key value of '7' is referenced as node '7'.

Figure 3 shows a uni-dimensional standard skip graph of eight nodes based on the sensor network in Figure 1. Each node contains a single-dimensional piece of data. The keys are [2, 3, 4, 7, 10, 12, 17, 20]. The randomly computed membership vectors associated with these nodes keys are [$010_2$, $111_2$, $001_2$, $110_2$, $010_2$, $001_2$, $101_2$, $010_2$], respectively.

In Figure 3, $L_0$ consists of a single linked list composed of all of the nodes in the network, and is sorted by node key value. The next highest level, $L_1$, consists of two linked lists. Node '3', node '7', and node '17' are grouped into one list, while the remaining nodes are grouped into another list. Membership vector prefixes determine which group a particular node falls in.

As the level number increases, the length of the membership vector prefix that is used to group nodes also increases. In $L_1$ of Figure 3, the first bit of the membership vector prefix is used to determine which list a node belongs. The membership vectors for Node '3', node '7', and node '17' all begin with a '1', and so those nodes are grouped together. The membership vectors for the remaining nodes all begin with '0', and so those nodes are placed into the other sub-list. In $L_2$, the first two bits of the membership vector determine which sub-list a node belongs. Therefore, there are four sub-lists in $L_2$ corresponding with the membership vector prefixes [00, 01, 10, 11], respectively. If there was no node that had a given 2-bit membership vector prefix of any of the prefixes in this list, a sub-list would not exist on that level corresponding with that prefix.

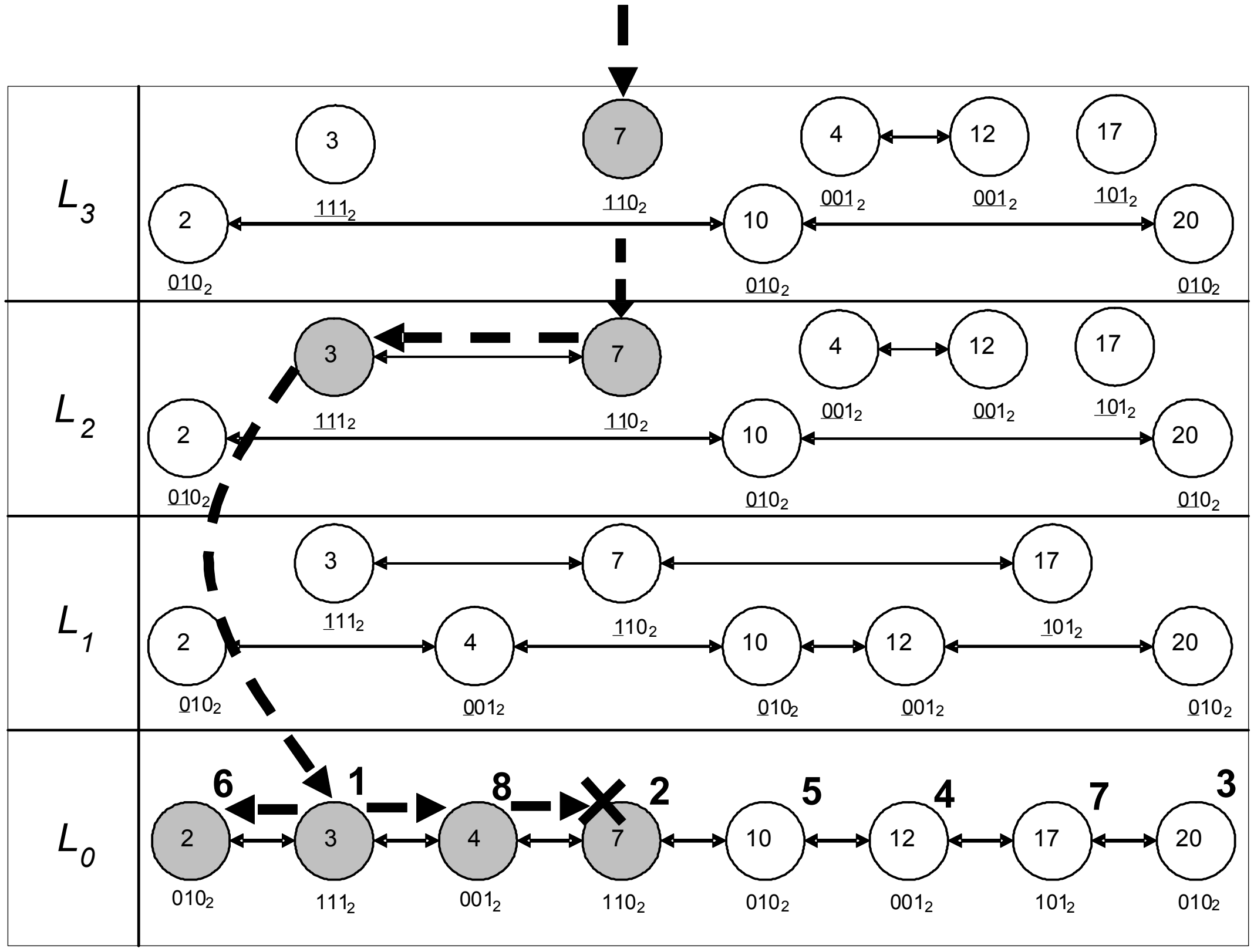

**Figure 3. Uni-dimensional standard skip graph**

### 3.2. Querying

Skip graphs are efficient structures for answering point and range queries. In a sensor network, a query may be injected into the network at any node. The query must propagate throughout the network and “find” all nodes according to the parameters of the query. In a range query, all nodes haveing key values in the specified range should be notified of the query.

To “find” the appropriate nodes, a search is performed for any node in the skip graph with a key value within range. Once found, the query drops to the lowest level $L_0$ and traverses that base list in both directions until it reaches key values that fall outside the desired range. At this point, the query can cease execution since the base list contains all nodes in the skip graph.

In Figure 3, a range query is injected at node ‘7’. This query is to contact all nodes with key values within the range of ‘2’ and ‘4’, inclusive. Since the query range falls below the value of the key at node ‘7’, and node ‘7’ has no references to any other nodes at $L_3$, the query must drop to level $L_2$. At this level, node ‘7’ has a reference to node ‘3’, i.e., a key value of ‘3’), and so the query transfers over to node ‘3’.

The key value at node ‘3’ is within the specified range, so the query drops down to the base level $L_0$ at node ‘3’ and advance in both directions, ensuring that all nodes within the query range ($2 \leq$ key $\leq 4$) are contacted. The query can stop (as indicated by the ‘X’ in $L_0$) when it finds a node in $L_0$ that has a key value outside of the range. In this case, that node happens to be the same node the query started with, node ‘7’.

## 4. Multi-dimensional standard skip graph

### 4.1 Construction

A multi-dimensional standard skip graph uses the same construction and querying process as the uni-dimensional standard skip graph, but we compute node keys from k-D data, such as a node’s geographic position. A node maps its k-D data (locally) to a 1-D key using a linearization method, such as *z*-ordering (cf. Section 2). A node’s membership vector is computed randomly, also locally. Figure 4 shows an example multi-dimensional standard skip graph.

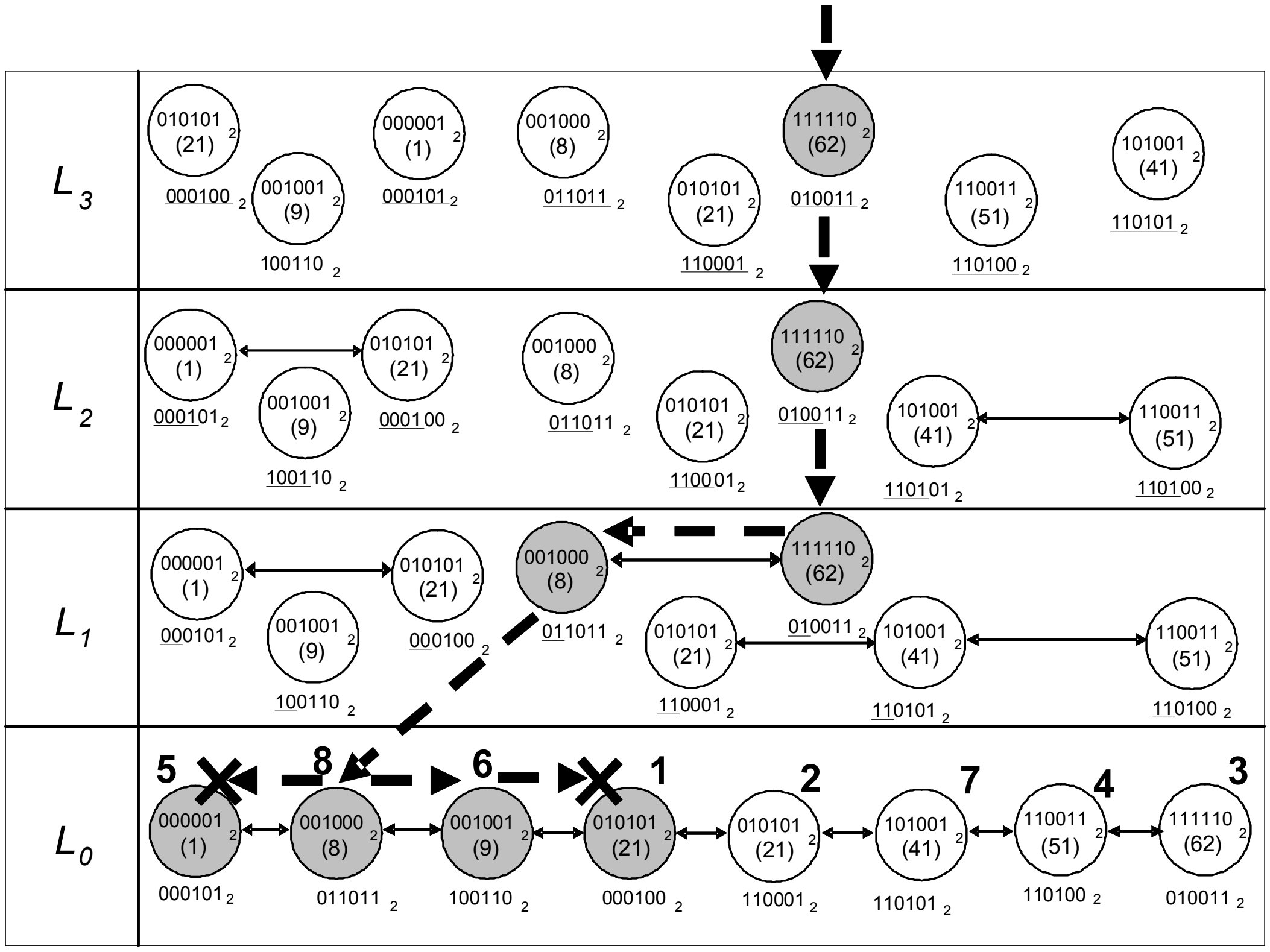

**Figure 4. Multi-dimensional standard skip graph**

The skip graph contains the same nodes, but the key values are deterministically computed from the $k$-D data—the $(x, y)$ coordinates of the nodes in Figure 1(b): [(0, 1), (2, 0), (2, 1), (0, 7), (0, 7), (6, 1), (5, 5), (7, 6)]. The corresponding $z$-order key values are [$000001_2$, $001000_2$, $001001_2$, $010101_2$, $010101_2$, $101001_2$, $110011_2$, $111110_2$], computed by using $z$-ordering. The numbers inside the parentheses of each node are the decimal equivalent of the node keys. The random membership vectors that we assigned to these nodes are [$000101_2$, $011011_2$, $100110_2$, $000100_2$, $110001_2$, $110101_2$, $110100_2$, $010011_2$], respectively.

### 4.2 Querying

Multi-Dimensional range querying is especially useful for locating a set of nodes within a certain geographic bound. The querying process for a multi-dimensional standard skip graph executes in a similar manner as a query in a uni-dimensional standard skip graph. For example, consider a geographical range query using the network illustrated in Figure 1. The range is a rectangle, bounded on the lower-left by (2, 0) and on the upper-right by (3, 1), that should return nodes {'6', '8'}. Since key values in a multi-dimensional standard skip graph are $z$-ordered linearizations of the node's geographic coordinates, the range query is first converted to a $z$-order query. The $z$-order of the first coordinate (2, 0) becomes ($001000_2$) and the $z$-order of the second coordinate (3, 1) becomes ($001011_2$). Thus, the query range becomes ($001000_2 – 001011_2$).

Figure 4 uses this range query of ($001000_2 – 001011_2$). The query is injected into the skip graph at node '62', and the process for executing the range query in a multi-dimensional standard skip graph is the same as described in Section 3.2. Once a node is found that is within the query range, the query drops to the base list $L_0$ and advances in both directions until nodes with key values outside of the query range are found. This is indicated in Figure 4 by the two 'X's. The two nodes found are nodes '6' and '8', which as previously mentioned, are the two nodes that fall within the specified query range.

## 5. Multi-dimensional inverted skip graph

### 5.1. Overview

Our proposed modification to the standard skip graph inverts the roles of deterministic key values and random membership vectors. In a multi-dimensional *inverted* skip graph, we compute node keys randomly at each node in the network, much like the random membership vector computations in the {uni, multi}-dimensional standard skip graphs.

Similarly, we compute node membership vectors deterministically from the $k$-D data being stored, akin to how we compute keys in {uni, multi}-dimensional standard skip graphs (cf. Sections 3 and 4). However, we still use $Z$-ordering to linearize the $k$-D data; the key difference is the $z$-order bit sequence is assigned to a node's membership vector, versus its key.

### 5.2. Construction

Figure 5 is a multi-dimensional inverted skip graph based on the network shown in Figure 1. The same eight nodes are used. Node key values are determined randomly; in Figure 5, they are [5, 13, 29, 37, 40, 63, 70, 89]. The membership vectors are then computed deterministically from the $z$-order mapping of a node's coordinates. The membership vectors, corresponding to the (random) node key values above are [$001001_2$, $010101_2$, $101001_2$, $010101_2$, $000001_2$, $110011_2$, $001000_2$, $111110_2$], respectively.

### 5.3. Querying

Range queries in a multi-dimensional inverted skip graph, are injected at any node as it is in the standard skip graphs. However, instead of the query starting at the topmost level, the query begins at the base level $L_0$ and traverses in both directions until it finds a node that has an $L_1$ membership vector prefix that matches the query's prefix for $L_1$. When the query finds that node, it advances to $L_1$ and traverses the list associated with the matching membership vector following a similar process in the base list only now the query is looking for a longer prefix to match. The query continues this process of longer prefix matching as it advances up the skip graph levels and terminates when one of three conditions is met.

The first condition is when the query's prefix fully matches the prefix of a membership vector on some level of the skip graph. This means that every node within that list should be notified of the query. The query progresses in both directions until reaching both ends of the list. For example, Figure 5 illustrates the execution of a query range of ($001000_2$ – $001011_2$). Essentially, this query returns all nodes that have a membership vector prefix of (0010) since that is the longest common prefix over the entire range. The query is injected into the skip graph at node '89' and progresses to the left in the base list $L_0$.

The query is then transferred to node '70', where it is determined the $L_1$ membership vector prefix (00) of node '70' matches the query (0010X) at that level. At this point, the query remains within node '70', but advances to $L_1$. The $L_2$ membership vector prefix (0010) of node '70' also matches that of the query (0010X) at that level. Thus, the query traverses the entire list at this level in which node '70' resides, because the query prefix of (0010) exactly matches. As Figure 5 shows, the two nodes returned in the query ('5' and '70'), are the two nodes that fall within the specified query range of ($001000_2$ – $001011_2$).

The second condition that stops the query is when the query advances to the highest level in the skip graph. At this point, the query prefix is compared with the longest possible membership vector prefix. The query must traverse this entire list, and at each node it compares the entire query prefix with the appropriate prefix length of the membership vector.

The third condition that stops the query occurs when the query traverses an entire list without finding any nodes in that list that have an $L_{i+1}$ (the next highest level) membership vector prefix that matches the query prefix for the level the query is currently in. If we modify the example in Figure 5 such that the query is (0011X), upon reaching $L_1$ the query would traverse through the entire list of node '70', node '40', and node '5', and not find any $L_2$ membership vector prefixes of (0011).

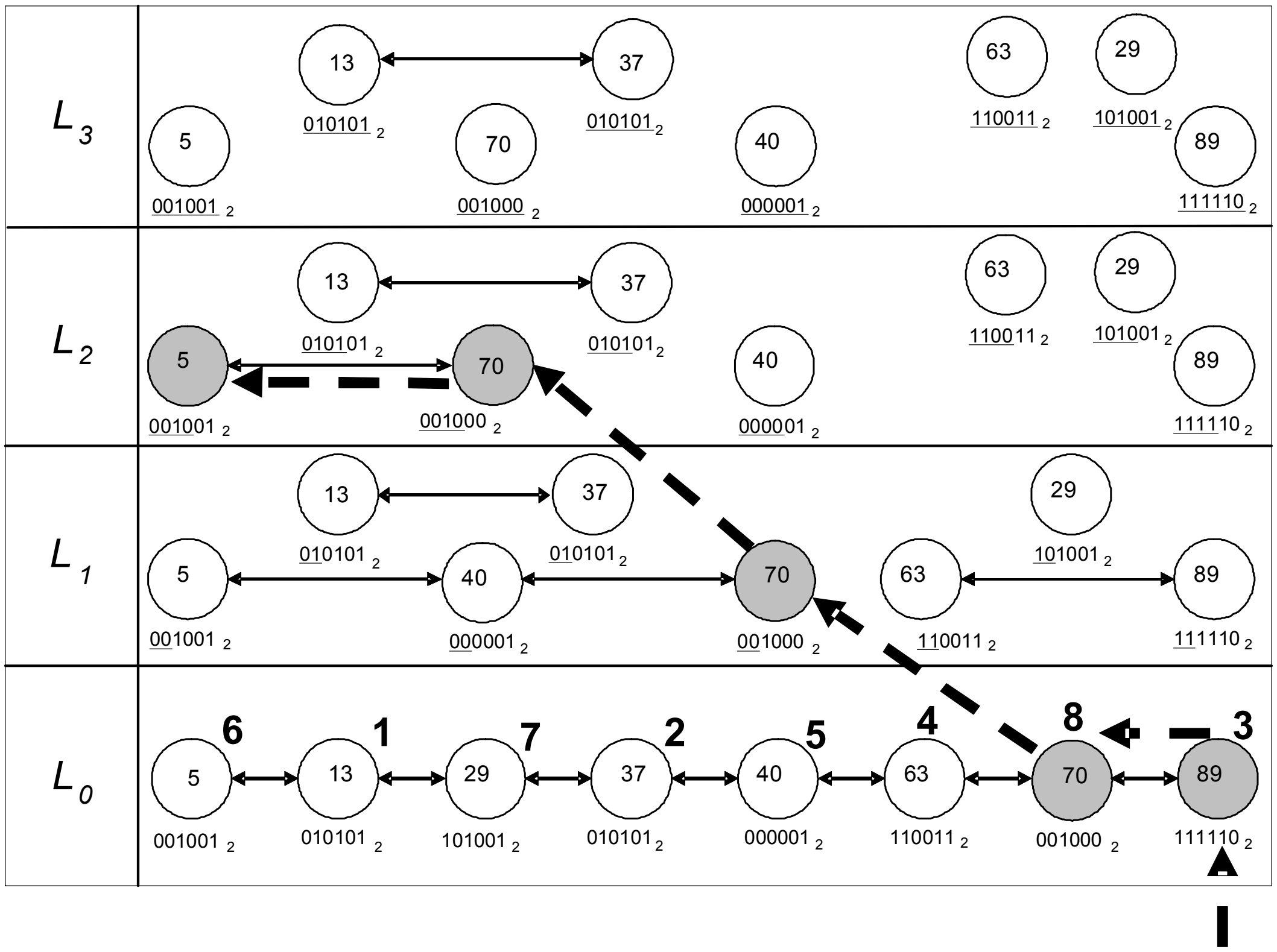

**Figure 5. Multi-dimensional inverted skip graph**

## 6. Conclusions

An extension to a standard skip graph is proposed. Standard skip graphs group nodes based on common prefixes of randomly computed membership vectors and sorted on deterministically computed key values. Our *k*-D inverted skip graph exchanges these roles by grouping nodes on common prefixes of deterministic membership vectors and sorts the nodes in each group on randomly computed keys. Our preliminary results indicate grouping nodes based on the *k*-D information being queried yields better performance in terms of total messages sent, if certain query conditions are met, especially in mobile environments.

## 7. Future work

Future work includes defining the insert, delete, and repair functions for inverted skip graphs, similar to how their standard skip graph definitions [1][6]. Simulation studies of queries executed in *k*-D standard and inverted skip graphs should be conducted to assess the number of messages needed to maintain them, respectively. Since network-layer routing costs are different as compared to skip graph routing costs, it would be useful to study the number of network-layer messages, hops, and geographic distance. We are particularly interested in assessing how the power consumption of maintaining the skip graph and its effects on a node's lifetime.

Mobility plays an important factor in sensor networks. Mobile nodes will ultimately break links between nodes in a skip graph due to communication failures or changes in geographic locations of the nodes, thus spawning similar changes in node keys or membership vectors. We are researching the effects of node mobility on both skip graph types, especially the messages needed to maintain the skip graph and how efficiently it responds to range queries in a 3-D mobile environment, e.g., a large UAV swarm. In a UAV swarm, we are primarily interested in how well a skip can be maintained and query response time, and are less concerned with power consumption.

## 8. Acknowledgments

We thank Gauri Shah for providing us source code to the standard skip graph [1]. We also thank Hanan Samet for providing us an advance copy of [11].